# Anisotropic exceptional points of arbitrary order

Yi-Xin Xiao[1,2], Zhao-Qing Zhang[2], Zhi Hong Hang[1,3], and C. T. Chan[2*]

[1]*School of Physical Science and Technology, Soochow University, Suzhou 215006, China*

[2]*Department of Physics, Hong Kong University of Science and Technology, Hong Kong*

[3]*Institute for Advanced Study, Soochow University, Suzhou 215006, China*

A pair of anisotropic exceptional points (EPs) of arbitrary order are found in a class of non-Hermitian random systems with asymmetric hoppings. Both eigenvalues and eigenvectors exhibit distinct behaviors when these anisotropic EPs are approached from two orthogonal directions in the parameter space. For an order-$N$ anisotropic EP, the critical exponents $\nu$ of phase rigidity are $(N-1)/2$ and $N-1$, respectively. These exponents are universal within the class. The order-$N$ anisotropic EPs split and trace out multiple ellipses of EPs of order 2 in the parameter space. For some particular configurations, all the EP ellipses coalesce and form a ring of EPs of order $N$. Crossover to the conventional order-$N$ EPs with $\nu = (N-1)/N$ is discussed.

*Introduction.*—There has been a surge of interest in the physics of non-Hermitian systems [1–48]. For classical wave systems, the attention has been focused on a broad range of systems involving material gain/loss or radiation leaking into open space. A prominent property of non-Hermitian systems is the existence of singularities called exceptional points (EPs), at which the Hamiltonian matrix becomes defective and multiple eigenstates coalesce [28–46,48–57]. EPs have been studied in various non-Hermitian systems due to their fascinating properties, which accounts for unusual transmission or reflection [9] and potential applications in lasing [15,19] and sensing [39,41]. Non-Hermiticity induces topological properties with new features that cannot be found in Hermitian systems [25,53,58–62].

The most common EPs are "order-2" EPs at which two eigenstates coalesce. An order-2 EP is usually associated with a square-root singularity in the eigenvalues, with $E \propto \pm\sqrt{\delta}$ where $\delta$ denotes a small deviation from the EP in the parameter space [54], and is associated with two Riemann sheets in the complex $\delta$-plane. It requires encircling two cycles around the order-2 EP ($\delta = 0$) to bring a state back to itself [3,28]. The coalescence of two eigenstates at the EP makes the right and left coalesced eigenvectors orthogonal, i.e., $\langle \psi_{EP}^L | \psi_{EP}^R \rangle = 0$, where the superscripts $L$ and $R$ signify left and right eigenvectors [55]. As a result, the so-called phase rigidity $\rho_m = \frac{|\langle \psi_m^L | \psi_m^R \rangle|}{\psi_m^R | \psi_m^R \rangle}$ of each eigenstate $m$ behaves in a power law $\rho_m \propto |\delta|^\nu$ at small $|\delta|$ with $\nu = 1/2$ [35,44,63]. The exponent $\nu$ measures the rate that the left and right eigenvectors turn orthogonal when an EP is approached. Phase rigidity is the inverse square-root of the Petermann factor which underlies the enhancement in intrinsic laser linewidth and spontaneous emission rate due to the non-orthogonality of laser modes [36,64–66]. The coalescence of multiple EPs can lead to a higher-order EP [36]. At an order-$N$ EP, the matrix is highly defective and $N$ eigenstates coalesce. Near an order-$N$ EP, the $N$ eigenvalues generally follow the $\sqrt[N]{\delta}$ singularity with one eigenvalue in each of the $N$ Riemann sheets in the complex $\delta$-plane and $N$ rounds of adiabatic encircling are needed to return to the initial sheet [35]. For an order-$N$ EP, the exponent of the phase rigidity becomes $\nu = \frac{N-1}{N}$ [30,67]. Higher-order EPs have been suggested to enhance sensing capabilities [39]. However, it is difficult to achieve a higher order EP as it requires some detailed tuning of multiple parameters.

Recently, order-2 anisotropic EPs are studied both theoretically and experimentally in different physical systems [35,44,60,56]. An anisotropic EP is associated with different singular behaviors when the EP is approached from two orthogonal directions in a two-parameter space, e.g., $(\xi, \eta)$-plane [44]. They are also called hybrid EPs [60]. Near an order-2 anisotropic EP ($\xi = \eta = 0$), two eigenvalues follow the behavior of $E \propto \pm\sqrt{\eta^2 + B\xi}$, where $B$ is a constant [44,60]. Along the line $\eta = 0$, the $\pm\sqrt{B\xi}$ singularity is followed. However, along the orthogonal direction, $\xi = 0$, two eigenvalues exhibit linear crossing behavior, namely $E \propto \pm\eta$, near the anisotropic EP [32,44,45]. The phase rigidity is also anisotropic bearing two different exponents: $1/2$ (along $\eta = 0$) and $1$ (along $\xi = 0$), i.e., $\rho \propto |\xi|^{1/2}$ and $\rho \propto |\epsilon|$, respectively.

In this work, we show that a class of non-Hermitian random systems with asymmetric hoppings can exhibit two anisotropic EPs of arbitrary high order. Near the order-$N$ anisotropic EP, the eigenvalues have the form $A_i\sqrt{\delta}$ and $B_i\epsilon$ in two orthogonal directions, respectively, where $A_i$ and $B_i$ are constants and $\delta$ and $\epsilon$ denote small deviations from the anisotropic EP. The phase rigidity shows a power-law behavior with two universal exponents, $(N-1)/2$ and $N-1$, which reduces to the exponents of $1/2$ and $1$ found recently for $N = 2$ anisotropic EPs. Interestingly, the two order-$N$ anisotropic EPs split and give rise to multiple elliptical trajectories of order-2 EPs in the parameter space, and these EP trajectories always merge at the two order-$N$ anisotropic EPs, which act as their common vertices. In a particular configuration, corresponding to the non-Hermitian Hamiltonian being linearly spanned by three spin-$j$ matrices, all the order-2 exceptional ellipses coalesce and form a ring of EPs of order $N = 2j + 1$, which is the generalization of the ring of order-2 EPs found recently in photonic crystal slabs [32] to arbitrary higher order.

*Two Anisotropic EPs of arbitrary order.*—To put our discussion in perspective, we start with a simple system that admits closed form solutions: an ordered asymmetric nearest-neighbor hopping model with $N$ sites [62,68]. The non-Hermitian Hamiltonian is a $N \times N$ tridiagonal matrix:

$$H = \begin{pmatrix} 0 & t_1 & 0 & 0 & 0 \\ t_2 & 0 & t_1 & 0 & 0 \\ 0 & t_2 & 0 & \ddots & 0 \\ 0 & 0 & \ddots & \ddots & t_1 \\ 0 & 0 & 0 & t_2 & 0 \end{pmatrix}, \tag{1}$$

where $t_{1,2} = t \mp \gamma$ with $t$ and $\gamma$ being real. The system is non-reciprocal if $\gamma \neq 0$. For convenience, we normalized the energy scale by setting $\gamma = 1$ in this work. The physical implementation of such non-reciprocal hoppings were discussed originally in the context of magnetic flux lines in type-II superconductors [68]. The non-reciprocal hoppings can in principle be realized using coupled ring resonators [69,70]. Alternatively, the asymmetric hopping model can be implemented with ultracold atoms in optical lattices [62] or electronic circuits [71]. The eigenvalues of $H$ have the square-root form:

$$E_p = 2\sqrt{t^2 - 1} \cos\left(\frac{p\pi}{N+1}\right), p = 1,2,\cdots,N. \tag{2}$$

There are two EPs at $t = \pm 1$, at which Eq. (1) reduces to a Jordan block form or its transpose, and all $N$ eigenstates coalesce to a single one with eigenvalue $E_{EP} = 0$, leading to an EP of order $N$. We now introduce diagonal entries (on-site terms) through a new parameter $\epsilon$ to the Hamiltonian. For example, we consider $H' = H + \epsilon s_z^j$, where $s_z^j$ is related to the $z$ component of the spin-$j$ operator by $s_z^j = S_z^j/j\hbar$ with $j = \frac{N-1}{2}$. Explicitly, we have $\left(S_z^j\right)_{pq} = (j + 1 - p)\hbar\delta_{pq}$, where $\delta_{pq}$ is the Kronecker delta function and $p,q = 1,2,\dots,N$. In the $(t,\epsilon)$-plane, $(\pm 1,0)$ are two EPs of order $N$ for any $N$. These two EPs exhibit anisotropic behaviors. Taking $j = 3/2$ as an example, the eigenvalues of the $4 \times 4$ Hamiltonian $H'$ have the form,

$$E = \pm \frac{\sqrt{2}}{6} \sqrt{A \pm \sqrt{B}}, \tag{3}$$

where $A = 27(t^2 - 1) + 10\epsilon^2$ and $B = 405(t^2 - 1)^2 + 432(t^2 - 1)\epsilon^2 + 64\epsilon^4$. They reduce to square-root forms with $E = \pm \frac{\sqrt{5}\pm 1}{2}\sqrt{t^2 - 1}$ when $\epsilon = 0$ and to linear forms with $E = \pm \epsilon$ or $E = \pm \frac{1}{3}\epsilon$ when $t = \pm 1$. Thus, the eigenvalues show different dispersion behaviors along different directions in the neighborhood of the EP, which bears the characteristic signature of anisotropic EPs. For general values $j$, the eigenvalues of $H'$ reduce to square-root forms given by Eq. (2) when $\epsilon = 0$, and linear forms with $E_p = \frac{j+1-p}{j}\epsilon$ when $t = \pm 1$, where $p = 1, 2, \ldots, N$. Thus $(t, \epsilon) = (\pm 1, 0)$ are two anisotropic EPs of any order $N$ for $H'$. Recent studies of anisotropic EPs involved only the coalescence of two states, i.e., $N = 2$ [44,60]. Here we have shown that a very simple Hamiltonian $H'$ carries anisotropic EPs of arbitrary high order, and the anisotropy can be attributed to the fact that perturbations introduced to different entries (diagonal or superdiagonal/subdiagonal elements) lead to different singular behaviors. To characterize eigenstates near an anisotropic EP, we calculate the phase rigidity near the anisotropic EP at $(1, 0)$ in the $(t, \epsilon)$-plane and find $\rho_i \propto |\delta|^{(N-1)/2}$ with $\delta = t - 1$ when $\epsilon = 0$, and $\rho_i \propto |\epsilon|^{N-1}$ when $\delta = 0$ [72]. The exponents of phase rigidity are respectively $(N-1)/2$ and $(N-1)$ in the two orthogonal directions, also exhibiting anisotropic behaviors.

*Universality.*—Surprising, the above EPs and their anisotropic behaviors are universal even in the presence of certain types of disorder. Let us consider the following Hamiltonian with disorder:

$$H_{dis} = \begin{pmatrix} c_1\epsilon & a_1 t_1 & 0 & 0 & 0 \\ b_1 t_2 & c_2\epsilon & a_2 t_1 & 0 & 0 \\ 0 & b_2 t_2 & c_3\epsilon & \ddots & 0 \\ 0 & 0 & \ddots & \ddots & a_{N-1} t_1 \\ 0 & 0 & 0 & b_{N-1} t_2 & c_N \epsilon \end{pmatrix}, \tag{4}$$

where $a_i, b_i, c_i$ are arbitrary real numbers with $a_i, b_i > 0$, and $t_{1,2} = t \mp 1$. The Hamiltonian $H_{dis}$ in Eq. (4) describes a system with not only asymmetric but also random hoppings. When $a_i = b_i = 1$ and $c_i = (j + 1 - i)/j$, the Hamiltonian $H_{dis}$ reduces to $H' = H + \epsilon s_z^j$ discussed earlier where $j$ is the spin index. It turns out that the points $(\pm 1, 0)$ remain two anisotropic EPs in the presence of randomness. It can be shown rigorously that all eigenvalues take the square-root form, namely $E_i = \pm D_i \sqrt{t_1 t_2}$ when $\epsilon = 0$, where $D_i$ are constants [72]. Near each anisotropic EP, say $(1, 0)$, the eigenvalues of $H_{dis}$ show anisotropic behaviors: $E_i \approx \pm D_i\sqrt{2\delta}$ with $\delta = t - 1$ along $\epsilon = 0$, and $E_i = c_i \epsilon$ along $t = 1$. The critical exponents of the phase rigidity are respectively $(N-1)/2$ and $N-1$ for all eigenstates, independent of the values of $a_i, b_i, c_i$ [72].

For illustration, we consider three disordered systems of different dimensions: $N = 2, 3, 4$. One eigenstate is taken as a representative state to calculate the phase rigidity for each system. The phase rigidity for the three systems is shown in Fig. 1(a) and (b) for the two orthogonal trajectories in the parameter space. Both trajectories go through the anisotropic EP at $(t, \epsilon) = (1,0)$. The values of $a_i, b_i, c_i$ for each case are given in the supplemental material [72]. Along $\epsilon = 0$ trajectory, the critical exponents of phase rigidity near the anisotropic EP are $1/2, 1, 3/2$ for $N = 2, 3, 4$, respectively, conforming to the $(N-1)/2$ rule. Along the other trajectory ($t = 1$), the critical exponents are $1, 2, 3$, respectively, agreeing with the $N - 1$ rule.

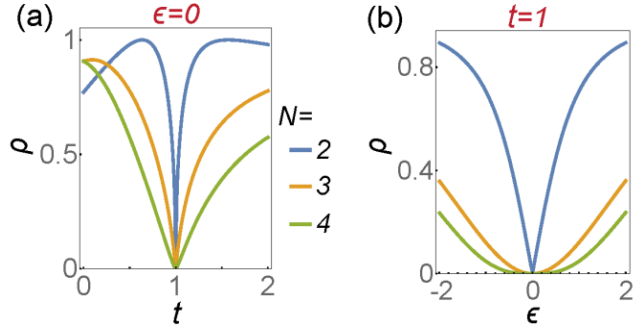

Fig. 1. Phase rigidity for three disordered systems $H_{dis}$ of different sizes, $N = 2, 3, 4$, along two orthogonal trajectories in $(t, \epsilon)$-plane: (a) $\epsilon = 0$; (b) $t = 1$.

Intriguingly, we find that the two order-$N$ anisotropic EPs at $(t, \epsilon) = (\pm 1, 0)$ split and trace out multiple elliptical trajectories formed by order-2 EPs in the form of $t^2 + \alpha_i \epsilon^2 = 1$ in the $(t, \epsilon)$-plane, with the points $(\pm 1, 0)$ being the common vertices of the ellipses, where $\alpha_i$ are some positive numbers depending on $a_i, b_i, c_i$. For the case of $N = 2$, the coalescence of two eigenvalues occurs on the ellipse $t^2 + \frac{(c_1 - c_2)^2}{4 a_1 b_1} \epsilon^2 = 1$, on which $E = \beta \epsilon$ with $\beta = (c_1 + c_2)/2$. For $N = 3$, it can be analytically shown that there always exists an ellipse of order-2 EPs in the form of $t^2 + \alpha \epsilon^2 = 1$, on which the coalesced eigenvalue is also proportional to $\epsilon$, i.e., $E = \beta\epsilon$. Here both $\alpha$ and $\beta$ depend on the values of all $a_i, b_i$ and $c_i$ [72]. Analytically we found $\alpha = -\frac{(c_1-\beta)(c_2-\beta)(c_3-\beta)}{a_1b_1(c_3-\beta)+a_2b_2(c_1-\beta)}$ for $N = 3$ scenario [72]. Here we conjecture that the proportionality between coalesced eigenvalue on each ellipse and $\epsilon$ holds for any $N$. With the ansatz $E = \beta\epsilon$, it can be argued from the characteristic equation $f_N(E) \equiv \det(H_{dis} - EI) = 0$ that the relation $t_1 t_2 = -\alpha\epsilon^2$ holds with $\alpha$ to be determined. Furthermore, at an EP, where two eigenvalues coalesce, we have $f'_N(E) = 0$. Thus, we can determine several solutions of $\beta_i$ and $\alpha_i$ by solving $f_N(E) = 0$ and $f'_N(E) = 0$, noting $\alpha_i = -t_1 t_2/\epsilon^2$ should be real. Therefore the order-2 EPs ellipses are $t^2 + \alpha_i\epsilon^2 = 1$ [72]. Here we would like to point out that all points on each ellipse are also anisotropic EPs [72].

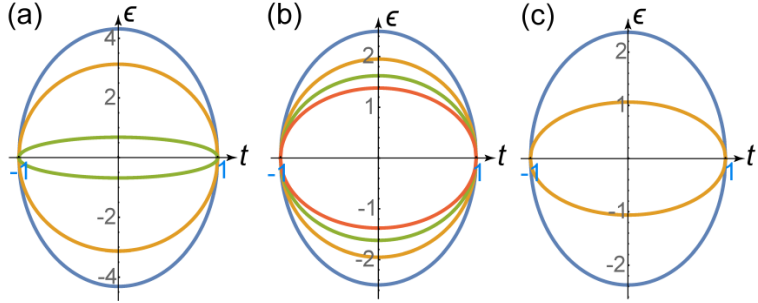

Fig. 2. Order-2 EPs form ellipses in the $(t, \epsilon)$-plane. (a) An $N = 6$ system. (b) An $N = 8$ system. The values of $a_i, b_i$ and $c_i$ are randomly chosen in (a) and (b) [72]. (c) The $N = 4$ ordered system $H' = H + \epsilon s_z^j$ with $j = 3/2$.

We found $N/2$ and $(N − 1)/2$ order-2 EPs ellipses, respectively, for even and odd $N$. For illustration, we show the numerically determined order-2 EP ellipses for $N = 6$ and $N = 8$ systems with random $a_i, b_i, c_i$ in Fig. 2 (a) and 2(b), respectively. There are three ellipses for $N = 6$ and four for $N = 8$. All ellipses have $(\pm 1, 0)$ as its vertices, giving two anisotropic EPs of order $N$. We note that closed form expressions for order-2 EPs ellipses can be found for the particular case $H' = H + \epsilon s_z^j$ with $j = 3/2$ considered previously, where $H$ is given by Eq. (1). Its eigenvalues are given by Eq. (3), which shows that the eigenvalues coalesce pairwise when $B = 0$. This condition gives two ellipses: $t^2 + \frac{8}{45}\epsilon^2 = 1$ and $t^2 + \frac{8}{9}\epsilon^2 = 1$ as depicted in Fig. 2(c). We emphasize that the ellipses of order-2 EPs are always pinned by the two order-$N$ EPs at $(\pm 1, 0)$, with their shapes determined by the specific values of $a_i, b_i, c_i$.

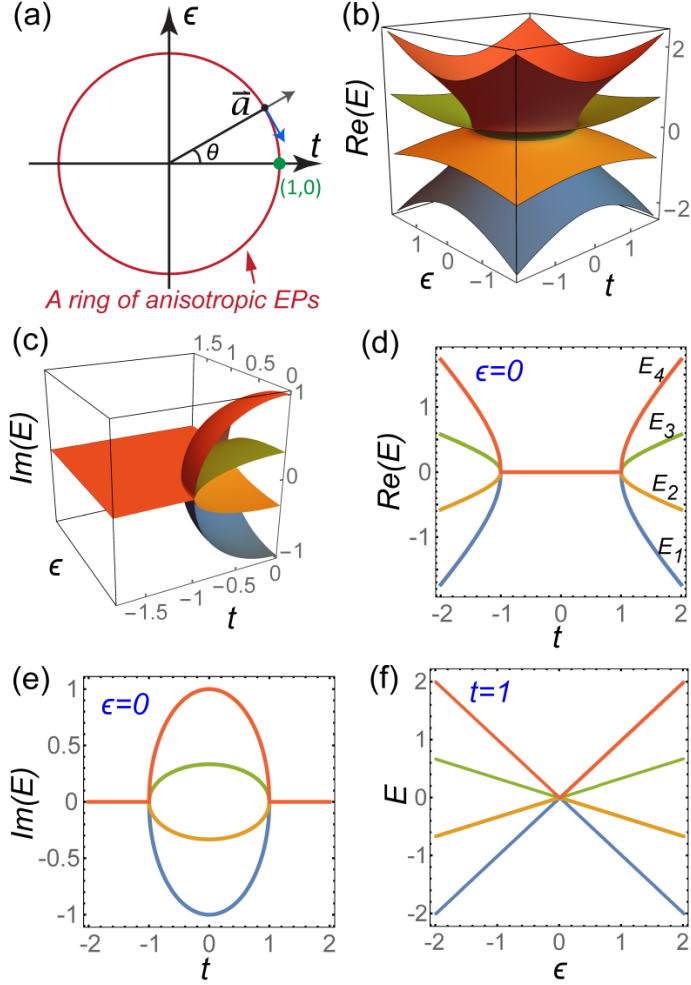

Fig. 3. (a) A ring of anisotropic exceptional points described by $t^2 + \epsilon^2 = 1$ for the Hamiltonian $H^j$. (b) The real part and (c) imaginary part of the eigen-energy surfaces for the case of $j = 3/2$. (d) and (e) The real and imaginary parts of the eigen-energies as functions of $t$ when fixing $\epsilon = 0$. (f) The eigen-energies as functions of $\epsilon$ when fixing $t = 1$.

*A ring of anisotropic EPs of arbitrary order.*—In the following, we show that when the coefficients $a_i, b_i, c_i$ corresponds to some linear combination of spin-$j$ matrices $s_\tau^j, \tau = x, y, z$, all ellipses of order-2 EPs will coincide and form a ring of order-$N$ EPs. In this case, the Hamiltonian takes the form

$$H^j = ts_x^j - is_y^j + \epsilon s_z^j, \tag{5}$$

where $s_\tau^j, \tau = x, y, z$ are spin-$j$ matrices defined by $s_\tau^j = S_\tau^j/j\hbar$, with $S_\tau^j$ being the spin-$j$ operators. The dimensions of $H^j$ matrix are $N \times N$, where $N = 2j + 1$. Explicitly, we have $s_x^j = \frac{s_+^j + s_-^j}{2}$ and $s_y^j = \frac{s_+^j - s_-^j}{2i}$, where $(s_+^j)_{pq} = \frac{\sqrt{(q-1)(2j+2-q)}}{j}\delta_{p,q-1}$ and $(s_-^j)_{pq} = \frac{\sqrt{(2j+1-q)q}}{j}\delta_{p,q+1}$ with $p, q = 1,2, \ldots, N$ [73]. We note that $H^j$ here and $H' = H + \epsilon s_z^j$ have the same form only when $j = \frac{1}{2}$ and $j = 1$, but not so for higher values of $j$.

Since Eq. (5) formally resembles the Hamiltonian of a spin in an artificial "magnetic field" $\vec{B} = (t, -i, \epsilon)$, we can always rotate the spin space to make the rotated $z'$-axis parallel to $\vec{B}$. Such a rotation transforms the Hamiltonian $H^j$ into $\mathcal{H}^j = R^{-1}H^j R = \sqrt{t^2 + \epsilon^2 - 1}\, s_z^j$ with $R = e^{-is_z^j j\phi_0}e^{-is_y^j j\theta_0}$, where $\theta_0 = \cos^{-1}\frac{\epsilon}{\sqrt{t^2+\epsilon^2-1}}$ and $\phi_0 = \cos^{-1}\frac{t}{\sqrt{t^2-1}}$ [74]. Thus, the eigenvalues of $\mathcal{H}^j$ can be derived immediately as

$$E_q = \frac{q-j-1}{j}\sqrt{t^2 + \epsilon^2 - 1}, \tag{6}$$

where $q = 1,2, \cdots, N$. Clearly the equation $t^2 + \epsilon^2 - 1 = 0$ represents a ring of order-$N$ EPs in the $(t, \epsilon)$-plane, which has unity radius. Each EP on the ring is actually an anisotropic EP [72]. To show this, we consider an arbitrary point $\vec{a} = (\cos\theta, \sin\theta)$ on the ring as depicted in Fig. 3(a). We introduce $\vec{d}_\perp = \delta(\cos\theta, \sin\theta)$ and $\vec{d}_\parallel = \delta(\sin\theta, -\cos\theta)$ with $0 < \delta \ll 1$ to represent small displacements from $\vec{a}$ in the radial and tangential directions of the ring, respectively, which are indicated by a gray and a blue arrow in Fig. 3(a). All $N$ eigenvalues exhibit a square-root form with $E_\perp \propto \sqrt{2\delta + \delta^2} \approx \sqrt{2\delta}$ at $\vec{a} + \vec{d}_\perp$. However, the $N$ eigenvalues are linear with respect to $\delta$, namely $E_\parallel \propto \delta$, at $\vec{a} + \vec{d}_\parallel$. Specifically, we show the real and imaginary parts of eigenvalue surfaces of the $j = 3/2$ case with $H^{3/2} = ts_x^{3/2} - is_y^{3/2} + \epsilon s_z^{3/2}$ in Figs. 3(b) and 3(c), where the anisotropic EP ring can be seen. Only a quarter of the ring is shown in Figs. 3(c). At point $\vec{a}$, all four eigenvalues of $H^{3/2}$ coalesce to $E = 0$, and all four right (left) eigenvectors also coalesce [72]. The square-root behaviors occurring at $(t, \epsilon) = (0, \pm 1)$ are shown in Figs. 3(d) and 3(e) when fixing $t = 0$ and varying $\epsilon$. The linear crossing occurring at $(0,1)$ is shown in Fig. 3(f). We note that the linear crossing point here is intrinsically different from the diabolic-point-like degeneracies found in Hermitian systems [75], because non-Hermiticity makes all $N$ eigenstates coalesce into a single one at the EP, where the phase rigidity vanishes. The phase rigidity is found to be $\rho_{i,\perp} \propto |\delta|^{3/2}$ at $\vec{a} + \vec{d}_\perp$ and $\rho_{i,\parallel} \propto |\delta|^3$ at $\vec{a} + \vec{d}_\parallel$ [72]. Thus, similar to the eigenvalues, phase rigidity also exhibits anisotropy: although the phase rigidity for both trajectories follows a power-law behavior, they vanish at the anisotropic EP with different critical exponents. The phase rigidity for all states are calculated numerically and shown in Fig. 4: 4(a) corresponds to Fig. 3(d,e), and 4(b) is for Fig. 3(f). We emphasize that the anisotropic behavior holds for $H^j$ of arbitrary dimensions. It can be shown rigorously that the phase rigidity behaves as $\rho_{i,\perp} \propto |\delta|^{(N-1)/2}$ and $\rho_{i,\parallel} \propto |\delta|^{N-1}$ in the radial and tangential directions, respectively, for any anisotropic EP on the ring [72].

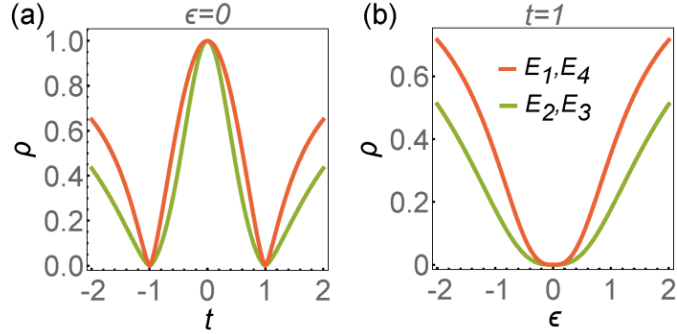

Fig. 4. The phase rigidity for spin-3/2 Hamiltonian $H^{3/2}$ along two trajectories: (a) the radial trajectory $\epsilon = 0$, and (b) the tangential trajectory $t = 1$ with respect to the EP at $(t, \epsilon) = (1,0)$ on the ring shown in Fig. 3(a). The critical exponents near the EP at $(t, \epsilon) = (1,0)$ are (a) $3/2$ and (b) $3$, respectively.

It is interesting to point out that $H^j = t s_x^j - i s_y^j + \epsilon s_z^j$ can be unitarily transformed to a formally *PT*-symmetric Hamiltonian with complex hoppings $H_{PT}^j = t s_x^j + \epsilon s_y^j + i s_z^j$ by a $\pi/2$ rotation of the spin space around the *x* axis. For the simplest case of $j = 1/2$, with the replacements $t \to k_x$ and $\epsilon \to k_y$, we end up with $H_{PT}^{1/2} = k_x s_x^{1/2} + k_y s_y^{1/2} + i s_z^{1/2}$, which is similar to the effective momentum space Hamiltonian of a non-Hermitian photonic crystal slab (periodic in xy plane) under symmetric gain and loss, where the ring of order-2 EPs are described by $k_x^2 + k_y^2 = 1$. Such an exceptional ring was experimentally observed [32,45]. EP rings of order 2 can spawn from Weyl points [53,57] when the system becomes non-Hermitian, which is also a special case of our considerations. Apart from engineering hoppings and onsite energies in the coupled ring resonator model according to Eq. (5), the ring of EPs of higher order can be realized by introducing non-Hermiticity into lattice systems bearing general spins, such as stacked triangular lattice layers [76].

*Discussion.—* We note that the relation $t_1 t_2 = -\alpha \epsilon^2$ for order-2 EP trajectories of $H_{dis}$ is very general. Instead of forming elliptical trajectories, the EPs will form a cluster of parabolas that can be parameterized by $t_1 = -\frac{\alpha}{t_2}\epsilon^2$ in the $(t_1, \epsilon)$-plane if we set $t_2$ (rather than the hopping difference $\gamma = (t_2 - t_1)/2$) as a constant.

We emphasize that the phase rigidity near the EP $(t, \epsilon) = (1,0)$ in the Hamiltonian in Eq. (1) behaves as $\rho_i \propto |\delta|^{(N-1)/2}$ with $\delta = t - 1$ when $\epsilon = 0$. The exponent $\nu = (N-1)/2$ in our class is very different from $\nu = \frac{(N-1)}{N}$ found in the conventional order-$N$ EP. The difference arises from the different singular behaviors in the eigenvalues near an EP. For Eq. (1), different pairs of the square-root singularities, i.e., $\pm a_i \sqrt{\delta}$, share the same EP, whereas for a conventional order-$N$ EP, all the $N$ eigenvalues come from one single $N$-th root singularity, i.e., $\sqrt[N]{\delta}$. Only two adiabatic encirclings are needed to bring a state back to itself in our case, instead of N encirclings needed for a usual order-$N$ EP. Also, a large value of exponent $\nu = (N-1)/2$, compared with $\nu = \frac{(N-1)}{N} < 1$, makes the characteristics of an EP, e.g., orthogonality of left and right eigenvectors, much easier to achieve.

We note that the conventional order-$N$ EP can also be easily obtained from Eq. (1) by setting $t_1 = 0$, i.e., $t = 1$, and at the same time insert a small perturbation $\delta$ to the upper-right corner entry of the Hamiltonian $H$. In fact, all the intermediate exponents between $(N-1)/2$ and $\frac{(N-1)}{N}$, namely $\frac{N-1}{n}$ with $2 < n < N$, can also be obtained by introducing perturbations to other entries [72].

This work is supported by Hong Kong Research Grants Council (grant AoE/P-02/12). Y.-X.X. is also supported by National Natural Science Foundation of China (No. 11847205) and China Postdoctoral Science Foundation (No. 2018M630597). This work is also funded by the Priority Academic Program Development (PAPD) of Jiangsu Higher Education Institutions.

*Corresponding author: phchan@ust.hk

**Supplemental material for "Anisotropic exceptional points of arbitrary order"**

Yi-Xin Xiao[1,2], Zhao-Qing Zhang[2], Zhi Hong Hang[1,3], and C. T. Chan[2*]

[1]*School of Physical Science and Technology, Soochow University, Suzhou 215006, China*
[2]*Department of Physics, Hong Kong University of Science and Technology, Hong Kong*
[3]*Institute for Advanced Study, Soochow University, Suzhou 215006, China*

## Table of Contents

# I. Anisotropic phase rigidity for ordered systems $H' = H + \epsilon s_z^j$

In the main text, we considered an asymmetric hopping model described by the matrix $H$ in Eq. (1), which is a $N \times N$ matrix. By adding a diagonal $\epsilon s_z^j$ term with $j = \frac{N-1}{2}$, we get

$$H' = H + \epsilon s_z^j = \begin{pmatrix} \epsilon & t_1 & 0 & 0 & 0 \\ t_2 & \frac{j-1}{j}\epsilon & t_1 & 0 & 0 \\ 0 & t_2 & \frac{j-2}{j}\epsilon & \ddots & 0 \\ 0 & 0 & \ddots & \ddots & t_1 \\ 0 & 0 & 0 & t_2 & -\epsilon \end{pmatrix}, \tag{S1}$$

where $t_1 = t - \gamma$ and $t_2 = t + \gamma$ and $\gamma = 1$. The points $(\pm 1, 0)$ are two AEPs of order $N$ in the $(t, \epsilon)$-plane. We have stated that the critical exponents of phase rigidity near the AEPs are $(N-1)/2$ and $N-1$, respectively, in two orthogonal directions for the $N \times N$ Hamiltonian $H'$. In the following, we give analytical derivations.

**(i) Derivation of exponent $(N-1)/2$ along $\epsilon = 0$ line**

Near the AEP at $(t, \epsilon) = (1,0)$ along the $\epsilon = 0$ line, we can write $t = 1 + \delta$, where $|\delta| = 0^+$ is a small number. Therefore $t_1 = t - 1 = \delta$ and $t_2 = t + 1 = 2 + \delta$. The $N \times N$ Hamiltonian $H' = H + \epsilon s_z^j$ becomes

$$H'(\delta) = \begin{pmatrix} 0 & \delta & 0 & 0 & 0 \\ 2+\delta & 0 & \delta & 0 & 0 \\ 0 & 2+\delta & 0 & \ddots & 0 \\ 0 & 0 & \ddots & \ddots & \delta \\ 0 & 0 & 0 & 2+\delta & 0 \end{pmatrix} \equiv \begin{pmatrix} 0 & \delta & 0 & 0 & 0 \\ G & 0 & \delta & 0 & 0 \\ 0 & G & 0 & \ddots & 0 \\ 0 & 0 & \ddots & \ddots & \delta \\ 0 & 0 & 0 & G & 0 \end{pmatrix}, \tag{S2}$$

where have taken $\epsilon = 0$ and defined $G \equiv 2 + \delta$ which is finite. We assume $\delta = 0^+$ for simplicity. The situation when $\delta = 0^-$ can be treated similarly.

The system has chiral symmetry: letting $\Gamma = diag\{1, -1, 1, -1, \cdots\}$, we have

$$\Gamma H' \Gamma = -H'.$$

Due to chiral symmetry, the eigen-spectrum of $H'(\delta)$ must be symmetric. That is, the eigenvalues come in pairs, i.e., $\pm E_1, \pm E_2, \cdots$, and there must be an $E = 0$ eigenvalue when $N$ is odd.

The eigenvalues are

$$E_m(\delta) = 2\sqrt{G\delta} \cos\left(\frac{m\pi}{N+1}\right), \tag{S3}$$

where $m = 1, 2, \cdots, N$. We use the notation $E_m(\delta) = \alpha_m \sqrt{\delta}$ for later convenience as long as $E_m(\delta) \neq 0$. The eigenstate corresponding to $E = 0$ when $N$ is odd will be discussed separately.

**(1) $E_m(\delta) \neq 0$ case**

The eigen-equation for an arbitrary $E_m(\delta) \neq 0$ eigenvalue can be written as

$$H'(\delta)|\psi_m^R(\delta)\rangle = E_m(\delta)|\psi_m^R(\delta)\rangle,$$

where $|\psi_m^R(\delta)\rangle = (x_1, x_2, \cdots, x_N)^T$ is the right eigenvector. Writing the eigen-equation explicitly into $N$ linear equations, and assuming $x_N = 1$, we immediately obtain

$$x_{N-1} = \frac{\alpha_m}{G}\sqrt{\delta} \propto \delta^{\frac{1}{2}}, \quad x_{N-2} = \frac{\alpha_m\sqrt{\delta}x_{N-1} - \delta x_N}{G} \propto \delta,$$

$$x_{N-3} = \frac{\alpha_m\sqrt{\delta}x_{N-2} - \delta x_{N-1}}{G} \propto \delta^{\frac{3}{2}}, \quad \cdots, \quad x_1 = \frac{\alpha_m\sqrt{\delta}x_2 - \delta x_3}{G} \propto \delta^{(N-1)/2}.$$

Thus, the right eigenvector $|\psi_m^R(\delta)\rangle$ can be expressed as

$$|\psi_m^R(\delta)\rangle = \left(A_{N-1}\delta^{\frac{N-1}{2}}, A_{N-2}\delta^{\frac{N-2}{2}}, \cdots, A_1\delta^{\frac{1}{2}}, A_0\right)^T, \tag{S4}$$

where the specific coefficients before $\delta^{i/2}, i = 1, 2 \cdots, N-1$ are represented by $A_i$ and $A_0 = 1$.

Similarly, the left eigenvector for $E_m(\delta) \neq 0$ is found to have the form

$$\langle\psi_m^L(\delta)| = \left(B_0, B_1\delta^{\frac{1}{2}}, \cdots, B_{N-2}\delta^{\frac{N-2}{2}}, B_{N-1}\delta^{\frac{N-1}{2}}\right), \tag{S5}$$

from $\langle\psi_m^L(\delta)|H'(\delta) = E_m(\delta)\langle\psi_m^L(\delta)|$, where the detailed coefficients are represented by $B_i$ and $B_0 = 1$.

In the main text, phase rigidity is defined as

$$\rho_m = \frac{1}{|\langle\tilde{\psi}_m^R|\tilde{\psi}_m^R\rangle|} = \frac{|\langle\psi_m^L|\psi_m^R\rangle|}{\langle\psi_m^R|\psi_m^R\rangle}, \tag{S6a}$$

where $|\tilde{\psi}_m^R\rangle = |\psi_m^R\rangle/\sqrt{|\langle\psi_m^L|\psi_m^R\rangle|}$ is biorthogonal normalized eigenvector. If we assume eigenvectors $|\phi_m^R\rangle$ and $\langle\phi_m^L|$ are normalized, namely $\langle\phi_m^R|\phi_m^R\rangle = 1$ and $\langle\phi_m^L|\phi_m^L\rangle = 1$, the phase rigidity can be expressed as

$$\rho_m = |\langle\phi_m^L|\phi_m^R\rangle|, \tag{S6b}$$

since the denominator in Eq. (S6a) is unity, i.e., $\langle\phi_m^R|\phi_m^R\rangle = 1$. As a consequence, the phase rigidity defined in Eq. (S6b) is always unity for eigenstates of a Hermitian system. And the phase rigidity obviously vanishes at an EP due to self-orthogonality, that is, $\langle\psi_{EP}^L|\psi_{EP}^R\rangle = 0$ [1,2]. As pointed out in the main text, phase rigidity is the inverse square-root of Petermann factor [3–5]. In G. C. New's definition of Petermann factor, both left and right eigenvectors are normalized [6], as we do for phase rigidity here.

The two eigenvectors $|\psi_m^R(\delta)\rangle$ and $\langle\psi_m^L(\delta)|$ in Eq. (S4) and (S5) are not normalized in the strict sense. Their normalization constants are

$$N_R = \langle\psi_m^R(\delta)|\psi_m^R(\delta)\rangle = \sum_{i=0}^{N-1}|A_i|^2\delta^i = 1 + \sum_{i=1}^{N-1}|A_i|^2\delta^i,$$

$$N_L = \langle\psi_m^L(\delta)|\psi_m^L(\delta)\rangle = \sum_{i=0}^{N-1}|B_i|^2\delta^i = 1 + \sum_{i=1}^{N-1}|B_i|^2\delta^i.$$

The normalized eigenvectors are $|\phi_m^R(\delta)\rangle = \frac{|\psi_m^R(\delta)\rangle}{\sqrt{N_R}}$ and $\langle\phi_m^L(\delta)| = \frac{\langle\psi_m^L(\delta)|}{\sqrt{N_L}}$. Thus the phase rigidity for $|\phi_m^R(\delta)\rangle$ is therefore

$$\rho_m(\delta) = |\langle \phi_m^L | \phi_m^R \rangle| = \frac{|\langle \psi_m^L | \psi_m^R \rangle|}{\sqrt{N_L N_R}} = \frac{\left| \sum_{i=0}^{N-1} A_i B_{N-1-i} \right| \delta^{\frac{N-1}{2}}}{\sqrt{(1 + \sum_{i=1}^{N-1} |A_i|^2 \delta^i)(1 + \sum_{i=1}^{N-1} |B_i|^2 \delta^i)}}.$$

Keeping only the dominant term, we have

$$\rho_m(\delta) \approx C \delta^{\frac{N-1}{2}} \propto \delta^{\frac{N-1}{2}}, \tag{S7}$$

where $C = \left| \sum_{i=0}^{N-1} A_i B_{N-1-i} \right|$ is nonzero in general.

Next we consider the other situation when $E_m(\delta) = 0$.

**(2) $E_m(\delta) = 0$ case**

When $N$ is odd, we must have a zero eigenvalue $E_p(\delta) = 2\sqrt{G\delta} \cos\left(\frac{p\pi}{N+1}\right) = 0$ with $p = \frac{N+1}{2}$ due to chiral symmetry. By similar procedures, the right and the left eigenvectors $|\psi_p^R(\delta)\rangle$ and $\langle \psi_p^L(\delta)|$ associated to $E_p(\delta)$ are found to have the following forms

$$|\psi_p^R(\delta)\rangle = \left( A'_{N-1} \delta^{\frac{N-1}{2}}, 0, A'_{N-3} \delta^{\frac{N-3}{2}}, \cdots, A'_4 \delta^2, 0, A'_2 \delta, 0, 1 \right)^T, \tag{S8}$$

$$\langle \psi_p^L(\delta)| = \left( 1, 0, B'_2 \delta, 0, B'_4 \delta^2, \cdots, B'_{N-3} \delta^{\frac{N-3}{2}}, 0, B'_{N-1} \delta^{\frac{N-1}{2}} \right), \tag{S9}$$

where $A'_i, B'_i$ are coefficients. Different from Eq. (S4) and (S5), we note that every second component of the eigenvectors is zero, which is a property dictated by chiral symmetry. Immediately, we have

$$\rho_p(\delta) \propto \delta^{\frac{N-1}{2}}. \tag{S10}$$

Thus far, we have analytically shown that the phase rigidity for all eigenstates near the AEP is proportional to $\delta^{\frac{N-1}{2}}$ for the $N \times N$ Hamiltonian $H'$ along the $\epsilon = 0$ line. Next we derive the exponent $N - 1$ for phase rigidity along $t = 1$ line.

**(ii) Derivation of the exponent $N - 1$ along $t = 1$ line**

In the other direction when $\epsilon$ is varied and $t = 1$ is fixed, the critical exponent will become $N - 1$ instead of $(N - 1)/2$, as will be shown in the following.

With $t = 1$ fixed, the Hamiltonian $H' = H + \epsilon s_z^j$ become

$$H'(\epsilon) = \begin{pmatrix} \epsilon & 0 & 0 & 0 & 0 \\ 2 & \epsilon \frac{j-1}{j} & 0 & 0 & 0 \\ 0 & 2 & \epsilon \frac{j-2}{j} & \ddots & 0 \\ 0 & 0 & \ddots & \ddots & 0 \\ 0 & 0 & 0 & 2 & -\epsilon \end{pmatrix}. \tag{S11}$$

The eigenvalues of $H'(\epsilon)$ are $E_m(\epsilon) = \frac{m}{j}\epsilon, m = -j, -j+1, \cdots, j$. Following the similar procedures as the previous subsection, we can write the eigen-equation $H'(\epsilon)|\psi_m^R(\epsilon)\rangle = E_m(\epsilon)|\psi_m^R(\epsilon)\rangle$ explicitly and find that the right eigenvector takes the following form

$$|\psi_m^R(\epsilon)\rangle = (C_{N-1}\epsilon^{N-1}, C_{N-2}\epsilon^{N-2}, \cdots, C_1\epsilon, 1)^T, \tag{S12}$$

where $C_i$'s are constant coefficients. Similarly we can get the left eigenvector

$$\langle\psi_m^L(\epsilon)| = (1, D_1\epsilon, \cdots, D_{N-2}\epsilon^{N-2}, D_{N-1}\epsilon^{N-1})^T, \tag{S13}$$

where the coefficients $D_i$ may take zero values depending on the eigenvalue $E_m(\epsilon)$. The normalization constants are

$$N_R = \langle\psi_m^R(\epsilon)|\psi_m^R(\epsilon)\rangle = 1 + \sum_{i=1}^{N-1}|C_i|^2\epsilon^{2i}$$

$$N_L = \langle\psi_m^L(\epsilon)|\psi_m^L(\epsilon)\rangle = 1 + \sum_{i=1}^{N-1}|D_i|^2\epsilon^{2i}$$

The normalized eigenvectors are $|\phi_m^R(\epsilon)\rangle = \frac{1}{\sqrt{N_R}}|\psi_m^R(\epsilon)\rangle$ and $\langle\phi_m^L(\epsilon)| = \frac{1}{\sqrt{N_L}}\langle\psi_m^L(\epsilon)|$. Considering $|\epsilon| \ll 1$, the phase rigidity is

$$\rho_m(\epsilon) = |\langle\phi_m^L|\phi_m^R\rangle| = \frac{\sum_{i=0}^{N-1} C_i D_{N-1-i}\epsilon^{N-1}}{\sqrt{N_L N_R}} \propto \epsilon^{N-1}, \tag{S14}$$

where we used $C_0 = D_0 = 1$.

Up to now, we have demonstrated that the critical exponents of phase rigidity at the anisotropic EP are $(N-1)/2$ and $N-1$, respectively, in the two orthogonal directions in the parameter space.

## II. Eigenvalues of $H_{dis}$ have the square-root form when $\epsilon = 0$

Just like the uniform asymmetric hopping Hamiltonian $H$ in Eq. (1), the eigenvalues of the $N \times N$ Hamiltonian $H_{dis}$ in Eq. (4) will also take square-root form,

$$E_j = \alpha_j\sqrt{t_1 t_2}, \tag{S15}$$

where $j = 1, 2, \cdots, N$, when $\epsilon = 0$. In the following we substantiate the claim using mathematical induction. We use the notation $\mathcal{H}_{dis}^{N \times N} = H_{dis}^{N \times N}(\epsilon = 0)$.

For $\mathcal{H}_{dis}^{2\times 2} = \begin{pmatrix} 0 & a_1 t_1 \\ b_1 t_2 & 0 \end{pmatrix}$, the eigenvalues obviously take square-root form, i.e., $E_{1,2} = \pm\sqrt{a_1 b_1}\sqrt{t_1 t_2}$. Similarly, the eigenvalues of

$$\mathcal{H}_{dis}^{3\times 3} = \begin{pmatrix} 0 & a_1 t_1 & 0 \\ b_1 t_2 & 0 & a_2 t_1 \\ 0 & b_2 t_2 & 0 \end{pmatrix}, \tag{S16}$$

also take square-root form $E_{1,3} = \pm\sqrt{(a_1 b_1 + a_2 b_2)} \times \sqrt{t_1 t_2}$ and $E_2 = 0$.

We denote the characteristic polynomial of a general $\mathcal{H}_{dis}^{N \times N}$ by

$$f_N(E) = \det(\mathcal{H}_{dis}^{N \times N} - EI), \tag{S17}$$

where $I$ denotes identity matrix. By definition, we know that $f_N(E)$ can be expressed in terms of $f_{N-1}(E)$ and $f_{N-2}(E)$, which are characteristic polynomials of $\mathcal{H}_{dis}^{(N-1)\times(N-1)}$ and $\mathcal{H}_{dis}^{(N-2)\times(N-2)}$, respectively. That is

$$f_N(E) = -E f_{N-1}(E) - a_{N-1} b_{N-1} t_1 t_2 f_{N-2}(E). \tag{S18}$$

We assume that all eigenvalues of $\mathcal{H}_{dis}^{(N-2)\times(N-2)}$ and $\mathcal{H}_{dis}^{(N-1)\times(N-1)}$ take square-root form, which are $E_j = \mu_j \sqrt{t_1 t_2}$ and $E_j = \nu_j \sqrt{t_1 t_2}$ respectively, where $\mu_j$ and $\nu_j$ are constants. Therefore, we can write

$$f_{N-2}(E) = (-1)^{N-2} \prod_{i=1}^{N-2} (E - \mu_i \sqrt{t_1 t_2}) \tag{S19}$$

and

$$f_{N-1}(E) = (-1)^{N-1} \prod_{i=1}^{N-1} (E - \nu_i \sqrt{t_1 t_2}). \tag{S20}$$

Consequently, we can write

$$f_N(E) = (-1)^{N-2}[E \prod_{i=1}^{N-1}(E - \nu_i\sqrt{t_1 t_2}) - a_{N-1} b_{N-1} t_1 t_2 \prod_{i=1}^{N-2}(E - \mu_i\sqrt{t_1 t_2})]. \tag{S21}$$

We define $\sigma = E/\sqrt{t_1 t_2}$. Then Eq. (S21) can be rewritten as

$$f_N(E) = (-1)^{N-2}(\sqrt{t_1 t_2})^N[\sigma \prod_{i=1}^{N-1}(\sigma - \nu_i) - a_{N-1} b_{N-1} \prod_{i=1}^{N-2}(\sigma - \mu_i)]. \tag{S22}$$

The characteristic equation $f_N(E) = 0$ gives us

$$\sigma \prod_{i=1}^{N-1}(\sigma - \nu_i) - a_{N-1} b_{N-1} \prod_{i=1}^{N-2}(\sigma - \mu_i) = 0,$$

the roots $\sigma_i, i = 1,2,\cdots,N$ of which must be some numbers noting that $\mu_i$ and $\nu_i$ are constants. The roots of $f_N(E)$ are therefore

$$E_i = \sigma_i \sqrt{t_1 t_2}, \quad i = 1,2,\cdots,N. \tag{S23}$$

So all eigenvalues of a general $\mathcal{H}_{dis}^{N\times N}$ take the square-root form $E_i = \sigma_i \sqrt{t_1 t_2}$.

To mention, it is obvious that the eigenvalues of $H_{dis}$ take the form $E_i = c_i \epsilon$ when $t = 1$ is fixed so that $t_1 = t - 1 = 0$.

### III. Anisotropic phase rigidity for $H_{dis}$

We have stated in the main text that the critical exponents of phase rigidity, i.e., $(N-1)/2$ and $N-1$, in two orthogonal directions are universal, in the sense that they are the same independent of the random choices of the values of

$a_i, b_i, c_i$ in $H_{dis}$ in Eq. (4). The exponents $(N-1)/2$ and $N-1$ in the presence of disorder can be similarly derived as we have done for the uniform hopping scenario in Section II.

In Section III, we have proved that the eigenvalues are $E_i = \mu_i \sqrt{t_1 t_2}$ when fixing $\epsilon = 0$, and are $E_i = c_i \epsilon$ when fixing $t = 1$. Since the eigenvalues take the same forms as those of the uniform hopping model $H'$, we can follow the similar procedures in Section II to derive the analytic form of the phase rigidity. That is, we write the eigen-equation for the right and left eigenvectors explicitly and get the compact form of the eigenvectors near the anisotropic EP.

When $\epsilon = 0$, the right and left eigenvectors for $H_{dis}(\delta)$, where $\delta = t - 1$, take the same forms as Eq. (S4) and (S5), which can be straightforwardly obtained following procedures in Section II. Therefore the phase rigidity is given by Eq. (S7), namely

$$\rho(\delta) \propto \delta^{\frac{N-1}{2}}. \tag{S24}$$

When $t = 1$, the right and left eigenvectors for $H_{dis}(\epsilon)$ are given by Eq. (S8) and (S9). And the phase rigidity is given by Eq. (S10),

$$\rho(\epsilon) \propto \epsilon^{N-1}. \tag{S25}$$

Therefore, it is a universal fact that the critical exponents of the phase rigidity at the anisotropic EP at $(1,0)$ in the $t$-$\epsilon$ plane are $(N-1)/2$ and $N-1$, respectively. The conclusion does depend on the random coefficients $a_i, b_i$ and $c_i$ in Eq. (4).

## IV. Parameters used in Fig. 1 and Fig. 2

### (i) Fig. 1 parameters

The disordered $N \times N$ Hamiltonian $H_{dis}$ in Eq. (4) are specified by randomly chosen coefficients $a_1, a_2, \cdots, a_{N-1}$ and $b_1, b_2, \cdots, b_{N-1}$ and $c_1, c_2, \cdots, c_N$. In Fig. 1, coefficients are taken randomly picked in the uniform distribution $a_i, b_i, c_i \in [0.1, 2.0]$. We note, without loss of generality, the parameters for the systems with $N = 2$ and $N = 3$ are taken from those generated randomly for the system with $N = 4$, as given below.

Table 1. $a_i, b_i, c_i$ in Fig. 1.

|  | $a_1$ | $a_2$ | $a_3$ | $b_1$ | $b_2$ | $b_3$ | $c_1$ | $c_2$ | $c_3$ | $c_4$ |
|---|---|---|---|---|---|---|---|---|---|---|
| $N=4$ | 0.9 | 0.7 | 0.4 | 0.2 | 1.1 | 0.8 | 1.2 | 0.8 | 1.4 | 0.7 |
| $N=3$ | 0.9 | 0.7 |  | 0.2 | 1.1 |  | 1.2 | 0.8 | 1.4 |  |
| $N=2$ | 0.9 |  |  | 0.2 |  |  | 1.2 | 0.8 |  |  |

### (ii) Fig. 2 parameters

The coefficients $a_i, b_i$ and $c_i$ in Fig. 2(a) and 2(b) are taken randomly within the uniform distributions: $a_i, b_i \in [0.1,1]$ and $c_i \in [1,4]$, as detailed below.

Table 2. $a_i, b_i$ and $c_i$ for $N = 6$ system in Fig. 2(a).

| $a_1$ | $a_2$ | $a_3$ | $a_4$ | $a_5$ |  |
|---|---|---|---|---|---|
| 0.25 | 0.67 | 0.58 | 0.69 | 0.81 |  |
| $b_1$ | $b_2$ | $b_3$ | $b_4$ | $b_5$ |  |
| 0.74 | 0.57 | 0.59 | 0.48 | 0.6 |  |
| $c_1$ | $c_2$ | $c_3$ | $c_4$ | $c_5$ | $c_6$ |
| 3.31 | 2.05 | 1.63 | 2.72 | 2.37 | 2.97 |

Table 3. $a_i, b_i$ and $c_i$ for $N = 8$ system in Fig. 2(b).

| $a_1$ | $a_2$ | $a_3$ | $a_4$ | $a_5$ | $a_6$ | $a_7$ | |
|---|---|---|---|---|---|---|---|
| 0.7 | 0.16 | 0.85 | 0.84 | 0.79 | 0.79 | 0.59 | |
| $b_1$ | $b_2$ | $b_3$ | $b_4$ | $b_5$ | $b_6$ | $b_7$ | |
| 0.3 | 0.89 | 0.28 | 0.41 | 0.92 | 0.33 | 0.2 | |
| $c_1$ | $c_2$ | $c_3$ | $c_4$ | $c_5$ | $c_6$ | $c_7$ | $c_8$ |
| 3.27 | 3.86 | 1.58 | 3.23 | 2.68 | 1.09 | 2.21 | 1.88 |

## V. Exceptional ellipses: derivations

### (i) $N = 2$ case

The Hamiltonian has the form $H_{dis}^{N=2} = \begin{pmatrix} c_1\epsilon & a_1 t_1 \\ b_1 t_2 & c_2\epsilon \end{pmatrix}$. The characteristic polynomial is $f_2(E) = \det(H_{dis}^{N=2} - EI) = E^2 - (c_1\epsilon + c_2\epsilon)E + c_1 c_2 \epsilon^2 - a_1 b_1 t_1 t_2$ where $I$ means the identity matrix. Requiring $f_2(E) = 0$ gives us the eigenvalues

$$E = \frac{(c_1 + c_2)\epsilon \pm \sqrt{(c_1 - c_2)^2 \epsilon^2 + 4a_1 b_1 t_1 t_2}}{2}.$$

The degeneracy occurs when

$$(c_1 - c_2)^2 \epsilon^2 + 4a_1 b_1 t_1 t_2 = 0 \quad \Rightarrow \quad t^2 + \frac{(c_1 - c_2)^2}{4a_1 b_1} \epsilon^2 = 1, \quad (S26)$$

which give an EP2 ellipse explicitly. At the EP2 ellipse, we have

$$E = \frac{(c_1 + c_2)}{2} \epsilon, \quad (S27)$$

### (ii) $N = 3$ case

The Hamiltonian has the following form

$$H_{dis}^{N=3} = \begin{pmatrix} c_1\epsilon & a_1 t_1 & 0 \\ b_1 t_2 & c_2\epsilon & a_2 t_1 \\ 0 & b_2 t_2 & c_3\epsilon \end{pmatrix}.$$

The characteristic polynomial $f_3(E) = \det(H_{dis}^{N=3} - EI_3)$ and its derivative are

$$f_3(E) = (c_1\epsilon - E)(c_2\epsilon - E)(c_3\epsilon - E) - [a_1 b_1 (c_3\epsilon - E) + a_2 b_2 (c_1\epsilon - E)]t_1 t_2,$$
$$f_3'(E) = -(c_2\epsilon - E)(c_3\epsilon - E) - (c_1\epsilon - E)(c_3\epsilon - E) - (c_1\epsilon - E)(c_2\epsilon - E)$$
$$+ (a_1 b_1 + a_2 b_2) t_1 t_2.$$

At a two-fold degeneracy point, we have $f_3(E) = 0$ and $f_3'(E) = 0$. Combing the two equations and eliminating the $t_1 t_2$ term, we have

$$a_1 b_1 (c_1\epsilon + c_2\epsilon - 2E)(c_3\epsilon - E)^2 + a_2 b_2 (c_2\epsilon + c_3\epsilon - 2E)(c_1\epsilon - E)^2 = 0, \quad (S28)$$

which implies that the eigenvalues must be proportional to $\epsilon$. That is, the eigenvalues can be written in the form $E_i = \beta_i \epsilon$ when degeneracy occurs, where $\beta_i$ are constants.

With the proportional relation $E = \beta\epsilon$ substituted into the equation $f_3(E) = 0$ and assuming $\epsilon \neq 0$, we immediately obtain

$$t_1 t_2 = -\alpha\epsilon^2, \tag{S29}$$

where

$$\alpha = -\frac{(c_1-\beta)(c_2-\beta)(c_3-\beta)}{a_1 b_1(c_3-\beta) + a_2 b_2(c_1-\beta)}, \tag{S30}$$

which explains the EP2 ellipse observed in numerical results and hints that $\alpha > 0$. Conversely, the proportional relation $E = \beta\epsilon$ can be inferred from $f_3(E) = 0$ if we assume $t_1 t_2 = -\alpha\epsilon^2$.

It is obvious that we can get two equations for the two unknowns $\alpha$ and $\beta$, if we substitute $E = \beta\epsilon$ and $t_1 t_2 = -\alpha\epsilon^2$ into $f_3(E) = 0$ and $f_3'(E) = 0$. Solving the two equations and demanding $\alpha = -t_1 t_2/\epsilon^2$ to be real gives us solutions for $\alpha$ and $\beta$. With $\alpha$ determined, we know the equation of the EP2 ellipse, $t^2 + \alpha\epsilon^2 = 1$, which agrees with the diagonalization results.

*A few remarks:* Substituting $E = \beta\epsilon$ into Eq. (S28), we obtain

$$a_1 b_1 (c_3 - \beta)^2 (c_1 + c_2 - 2\beta) + a_2 b_2 (c_1 - \beta)^2 (c_2 + c_3 - 2\beta) = 0, \tag{S31}$$

which is the equation for $\beta$. When $a_2 b_2 = 0$ or $a_1 b_1 = 0$, $H_{dis}^{N=3}$ essentially reduces to $H_{dis}^{N=2}$ and Eq. (S31) gives us $\beta = \frac{c_1+c_2}{2}$ or $\beta = \frac{c_2+c_3}{2}$, which agrees with Eq. (S27).

**(iii) $N = 4$ case**

$$H_{dis}^{N=4} = \begin{pmatrix} c_1\epsilon & a_1 t_1 & 0 & 0 \\ b_1 t_2 & c_2\epsilon & a_2 t_1 & 0 \\ 0 & b_2 t_2 & c_3\epsilon & a_3 t_1 \\ 0 & 0 & b_3 t_2 & c_4\epsilon \end{pmatrix}$$

We observed that the proportionality $E = \beta\epsilon$ holds on the EP2 ellipse for the $N = 2$ case and $N = 3$ case. For $N = 4$, we take the ansatz that $E = \beta\epsilon$ on the EP2 ellipses, which is confirmed by numerical results. With $E = \beta\epsilon$, the relation $t_1 t_2 = -\alpha\epsilon^2$ can always be inferred from the characteristic equation $f_4(E) = 0$.

Solving $f_4(E) = 0$ and $f_4'(E) = 0$ with $E = \beta\epsilon$ and $t_1 t_2 = -\alpha\epsilon^2$ substituted, we can get $\alpha$ and $\beta$. Accordingly, two EP2 ellipses are determined: $t^2 + \alpha_1\epsilon^2 = 1$ and $t^2 + \alpha_2\epsilon^2 = 1$.

**(1) A concrete example for $N = 4$**

$$H_{dis}^{N=4} = \begin{pmatrix} 2.55\epsilon & 0.16 t_1 & 0 & 0 \\ 0.39 t_2 & 3.75\epsilon & 0.18 t_1 & 0 \\ 0 & 0.99 t_2 & 1.21\epsilon & 0.19 t_1 \\ 0 & 0 & 0.75 t_2 & 1.68\epsilon \end{pmatrix}$$

We note this $N = 4$ example system is the one used in the calculation for Fig. 2(a). Solving $f_4(E) = 0$ and $f_4'(E) = 0$ with $E = \beta\epsilon$ and $t_1 t_2 = -\alpha\epsilon^2$ substituted, we obtain six groups of solutions for $\alpha$ and $\beta$,

$$\alpha_1 = 0.34494, \quad \beta_1 = 1.45703,$$

$$\alpha_2 = 3.39979, \quad \beta_2 = 3.0338,$$
$$\alpha_{3,4} = -3.85 \pm 3.65\, i, \quad \beta_{3,4} = 2.28 \mp 0.046\, i,$$
$$\alpha_{5,6} = 13.13 \pm 5.76\, i, \quad \beta_{5,6} = 2.61 \mp 1.3\, i.$$

Since $t_1, t_2$ and $\epsilon$ are real parameters, we require $\alpha = -t_1 t_2/\epsilon^2$ to be real. Two EP2 ellipses are determined by
$$t^2 + \alpha_1 \epsilon^2 = 1 \quad \text{and} \quad t^2 + \alpha_2 \epsilon^2 = 1,$$
where $\alpha_1 = 0.34494$ and $\alpha_2 = 3.39979$. Therefore, the lengths of semi-major (or semi-minor) axes of the two ellipses along $\epsilon$ axis are
$$r_1 = \frac{1}{\sqrt{\alpha_1}} \approx 1.7 \quad \text{and} \quad r_2 = 1/\sqrt{\alpha_2} \approx 0.54,$$
which agrees with the results shown in Fig. 2(a). We note Fig. 2(a) is achieved by directly solving $f_4(E) = 0$ and $f_4'(E) = 0$ for real $\epsilon$ and $E$ for a given $t$, without the assumptions that $E = \beta \epsilon$ and $t_1 t_2 = -\alpha \epsilon^2$.

**(iv) Larger $N$ cases**

After dealing with $N = 4$, it is obvious that the ansatz $E = \beta \epsilon$ at the degeneracy and the consequent relation $t_1 t_2 = -\alpha \epsilon^2$ hold for a larger $N$. Similar procedures can be carried out to get solutions to $\alpha$ and $\beta$ and therefore the EP2 ellipses,
$$t^2 + \alpha_i \epsilon^2 = 1,$$
where $i = 1, 2, \cdots i_{\max}$, and $i_{max} = N/2$ for even $N$ and $i_{max} = (N-1)/2$ for odd $N$. The argument is as follows. For the real matrix $H_{dis}$, its eigenvalues should be either real or come in complex conjugate pairs. It has been rigorously proved that all eigenvalues of $H_{dis}$ take square-root form, i.e., $E \propto \sqrt{t_1 t_2} = \sqrt{t^2 - 1}$ when $\epsilon = 0$. Immediately, we know that all eigenvalues are purely real (imaginary) far away from (near) the origin on the $t$ axis. On the other hand, when $|\epsilon|$ goes to infinity, the hopping terms $a_i t_1, b_i t_2$ become insignificant and $H_{dis}$ tends towards Hermitian limit, therefore all eigenvalues should turn real. Therefore, going from the origin on the t-$\epsilon$ plane to infinity, all complex eigenvalues go real, as observed in numerical calculations. This implies the occurrence of EPs. It is reasonable that there are some EP boundaries enclosing the origin on the $t$-$\epsilon$ plane. For random scenarios, EP2 boundaries are most likely to occur. Therefore, $N/2$ EP2 ellipses are observed for even $N$ cases, and $(N-1)/2$ for odd $N$ cases. EP boundaries of higher order are only possible for specially chosen $a_i, b_i, c_i$, such as the exceptional ring cases.

## VI. All points on an EP2 ellipse are AEPs

The two EPNs at $(t, \epsilon) = (\pm 1, 0)$ split and trace out multiple EP2 ellipses on the $t$-$\epsilon$ plane, which has been analytically demonstrated in Section V. And we have shown that, for each spin form Hamiltonian $H^j$, all points on the exceptional ring are anisotropic exceptional points. We believe that all points on the EP2 ellipses in the case of $H_{dis}$ are also AEPs.

In the following, we show this point by considering a simple and concrete system $H_{dis}^{N=2} = \begin{pmatrix} c_1 \epsilon & a_1 t_1 \\ b_1 t_2 & c_2 \epsilon \end{pmatrix}$, where $a_1 = 1, b_1 = 3, c_1 = 8$ and $c_2 = 5$. In section V. (i), we have shown the EP2 ellipse is given by $t^2 + \alpha \epsilon^2 = 1$, where $\alpha = \frac{(c_1-c_2)^2}{4 a_1 b_1} = \frac{3}{4}$. A general point on the EP2 ellipse can be parameterized as $\vec{p} = (\cos\theta, \sin\theta/\sqrt{\alpha})$. And a small displacement from $\vec{p}$ in the radial direction can be expressed as
$$\vec{d}_\perp = \delta(\cos\theta, \sqrt{\alpha}\sin\theta). \tag{S32}$$
Similarly a small displacement from $\vec{p}$ in the tangential direction is
$$\vec{d}_\parallel = \delta(\sqrt{\alpha}\sin\theta, -\cos\theta). \tag{S33}$$
We assume $\delta \to 0^+$. Note the vectors denoted by parentheses in Eq. (S32) and (S33) are not normalized, which does not matter.

At the point $\vec{p} + \vec{d}_\perp = \left(\cos\theta, \frac{\sin\theta}{\sqrt{\alpha}}\right) + \delta(\cos\theta, \sqrt{\alpha}\sin\theta)$, we have $t = (1+\delta)\cos\theta$ and $\epsilon = \frac{1+\delta\alpha}{\sqrt{\alpha}}\sin\theta$. We found that the two eigenvalues satisfy $E - E_{EP} \propto \sqrt{\delta}$ for any $\theta$, where $E_{EP}$ is the eigenvalue at the exceptional point $\vec{p}$. And phase rigidity for two eigenstates is $\rho_\perp \propto \sqrt{\delta}$ for any $\theta$.

And at the point $\vec{p} + \vec{d}_\parallel = \left(\cos\theta, \frac{\sin\theta}{\sqrt{\alpha}}\right) + \delta(\sqrt{\alpha}\sin\theta, -\cos\theta)$, we have $t = \cos\theta + \delta\sqrt{\alpha}\sin\theta$ and $\epsilon = \frac{\sin\theta}{\sqrt{\alpha}} - \delta\cos\theta$. The two eigenvalues satisfy $E - E_{EP} \propto \delta$ for any $\theta$. And phase rigidity for two eigenstates is $\rho_\parallel \propto \delta$ for any $\theta$.

So we have shown all the points on the EP2 ellipse are also anisotropic EPs by exemplification.

## VII. One more example for EP2 ellipses

We consider the following Hamiltonian,
$$H = \begin{pmatrix} \epsilon & t_1 & 0 & 0 \\ t_2 & -\epsilon & t_1 & 0 \\ 0 & t_2 & \epsilon & t_1 \\ 0 & 0 & t_2 & -\epsilon \end{pmatrix}, \tag{S34}$$
where $t_1 = t - 1$ and $t_2 = t + 1$. The eigenvalues are
$$E = \pm \frac{(\sqrt{5} \pm 1)}{2} \sqrt{t^2 + \left(\frac{\sqrt{5} \mp 1}{2}\right)^2 \epsilon^2 - 1}.$$
Obviously there are two EP2 ellipses in the $t$-$\epsilon$ plane:

$$t^2 + \left(\frac{\sqrt{5}-1}{2}\right)^2 \epsilon^2 = 1,$$

$$t^2 + \left(\frac{\sqrt{5}+1}{2}\right)^2 \epsilon^2 = 1,$$

at which $E = 0$.

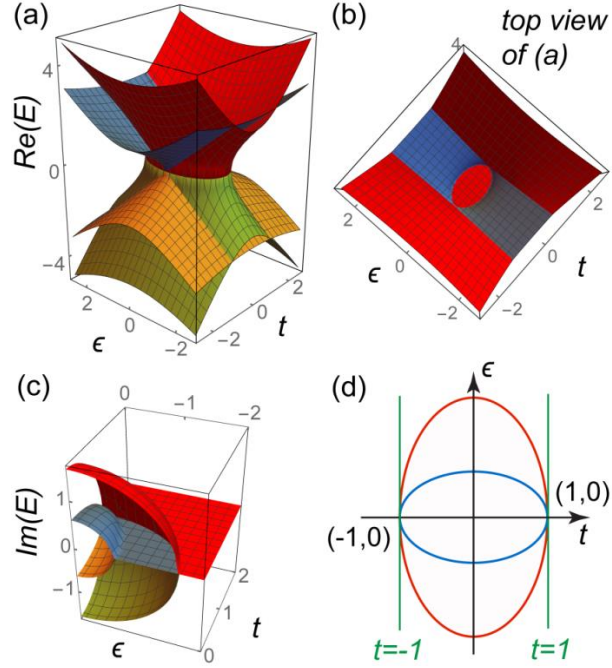

Fig. S1. The eigenvalues as a function of $t$ and $\epsilon$: (a) The real part; (b) is the top view of (a); (c) The imaginary part. (d) A view of the exceptional ellipses and two AEPs of order 4 in the $t$-$\epsilon$ plane.

Two EP2 ellipses meet at $(t, \epsilon) = (\pm 1, 0)$, giving rise to two EPs, each of order 4, as depicted in Fig. S1(d). The real and imaginary parts of eigenvalues are shown in Fig. S1(a) and S1(c), respectively, and Fig. S1(b) is the top view of Fig. S1(a). For clearness, only the quadrant with $t > 0, \epsilon < 0$ is shown for the imaginary part. The eigenvalues come in symmetric pairs. At each exceptional ellipse, one symmetric pair of eigenvalue surfaces intersects. Eigenvalue surfaces also cross at $t = \pm 1$, causing four exceptional lines of order 2. And the two fourth order EPs at $(\pm 1, 0)$ are the intersections of four surfaces. The larger ellipse $t^2 + \left(\frac{\sqrt{5}-1}{2}\right)^2 \epsilon^2 = 1$ is given by the intersection of the red and green surfaces, and the smaller one $t^2 + \left(\frac{\sqrt{5}+1}{2}\right)^2 \epsilon^2 = 1$ connects the yellow and navy surfaces. All degeneracy points corresponding to the ellipses and the two lines $t = \pm 1$ are EPs owing to the non-Hermiticity.

Just like the exceptional ring, the eigenvalues show square-root singularity behavior in the normal direction and disperse linearly in the tangential direction at each order-2 AEP on the ellipses. As expected from universality, the critical exponent of phase rigidity is 1/2 in the normal direction and is unity in the tangent direction. The AEPs of order 4 at $(\pm 1, 0)$ are also anticipated from previous analysis of the general Hamiltonian $H_{dis}$ in Eq. (4). Taking the one at $(1,0)$ as an example. Along $t = 1$, the tangential direction, the eigenvalues are purely real and form two pairs of EP2s. Therefore, the phase rigidity of each eigenstate vanishes in this direction. Along $\epsilon = 0$, however, all four eigenvalues show square-root type singularity near $(1,0)$, just like Fig. 2(d) and 2(e) for the exceptional ring scenario. The phase rigidity for the four eigenvalues is $\rho_1 = \rho_4 \approx \frac{1}{4}(5\sqrt{2} + \sqrt{10})\delta^{3/2}$ and $\rho_2 = \rho_3 \approx \frac{1}{4}(5\sqrt{2} - \sqrt{10})\delta^{3/2}$, where $\delta = t - 1$. The critical exponent 3/2 complies with universal results derived from $H_{dis}$. The phase rigidity vanishes along $t = 1$ since exceptions lines occur.

### VIII. Anisotropic phase rigidity: EP ring in "spin"-like Hamiltonians

**(i) Critical exponents of $(N-1)/2$ and $N-1$ at each EP on the ring**

We have shown that the critical exponents of phase rigidity are $(N-1)/2$ and $N-1$, respectively, near the two AEPs at $(\pm 1, 0)$ for the general disordered Hamiltonian $H_{dis}$. This is a universal result.

As a special case of $H_{dis}$, the "spin" Hamiltonian $H^j = ts_x^j - is_y^j + \epsilon s_z^j$ also has the two AEPs $(\pm 1, 0)$. The exponents of phase rigidity are also $(N-1)/2$ and $N-1$ along two directions. With a ring of AEPs, there are infinitely many other AEPs in addition to the two at $(\pm 1, 0)$. An arbitrary AEP on the ring can be parameterized as $(\cos\theta, \sin\theta)$. The Hamiltonian associated with $(t, \epsilon) = (\cos\theta, \sin\theta)$ can be unitarily transformed to the Hamiltonian associated with $(t, \epsilon) = (1,0)$ by a rotation around y axis.

According to the definition, a unitary transformation should not change the phase rigidity of an eigenstate. Therefore, we can conclude that the exponents of phase rigidity near an arbitrary AEP $(\cos\theta, \sin\theta)$ are $(N-1)/2$ and $N-1$, respectively, when approached in the radial and tangential directions.

**(ii) Spin-3/2 Hamiltonian as an example**

In the main text, we have stated that the phase rigidity for the spin-3/2 systems near the EPs on the ring conforms to the general rule demonstrated in Section IV. Here we show it by direct calculation.

We stated that $H^j$ and $H' = H + \epsilon s_z^j$ have the same form only when $j = \frac{1}{2}$ and $j = 1$, but not so for higher values of $j$. When we take $a_1 = a_3 = b_1 = b_3 = \frac{\sqrt{3}}{3}, a_2 = b_2 = \frac{2}{3}$ and $c_1 = -c_4 = 1, c_2 = -c_3 = 1/3$, the $4 \times 4$ Hamiltonian $H_{dis}$ reduces to the spin-3/2

Hamiltonian $H^{3/2} = ts_x^{3/2} - is_y^{3/2} + \epsilon s_z^{3/2}$, which is different from the uniform hopping model $H' = H + \epsilon s_z^{3/2}$.

An arbitrary point on the EP circle can be parameterized as $\vec{a} = (\cos\theta, \sin\theta)$. We can easily confirm that all four eigenvalues at $\vec{a}$ coalesce to $E = 0$, since $t^2 + \epsilon^2 - 1 = \cos^2\theta + \sin^2\theta - 1 = 0$. In addition, all the right eigenvectors coalesce to a single one, namely $\psi_{EP}^R = \left[\tan^3\left(\frac{\theta}{2}\right), \sqrt{3}\tan^2\left(\frac{\theta}{2}\right), \sqrt{3}\tan\left(\frac{\theta}{2}\right), 1\right]^T$. And all left eigenvectors coalesce to $\psi_{EP}^L = \left[-\cot^3\left(\frac{\theta}{2}\right), \sqrt{3}\cot^2\left(\frac{\theta}{2}\right), -\sqrt{3}\cot\left(\frac{\theta}{2}\right), 1\right]^T$.

We introduce two orthogonal displacement vectors $\vec{d}_\perp = \delta(\cos\theta, \sin\theta)$ and $\vec{d}_\parallel = \delta(\sin\theta, -\cos\theta)$ with $0 < \delta \ll 1$.

At the point $\vec{a} + \vec{d}_\perp = (1+\delta) \times (\cos\theta, \sin\theta)$, the Hamiltonian becomes
$$H_\perp^{3/2}(\theta) = (1+\delta)\cos\theta\, s_x^{3/2} - is_y^{3/2} + (1+\delta)\sin\theta\, s_z^{3/2}. \tag{S35}$$

Consistent with the general situation shown in the main text, the eigenvalues for $H_\perp^{3/2}(\theta)$ are $E_\perp = \frac{2}{3}m\sqrt{\delta^2 + 2\delta} \approx m\sqrt{2\delta}$ where $m = -\frac{3}{2}, -\frac{1}{2}, \frac{1}{2}, \frac{3}{2}$. The eigenvectors can be directly obtained by diagonalization. And we found the phase rigidities for $H_\perp^{3/2}(\theta)$ are

$$\rho_{1,\perp} = \rho_{4,\perp} = 2\sqrt{2}\delta^{3/2} + O(\delta^{5/2}), \tag{S36}$$
$$\rho_{2,\perp} = \rho_{3,\perp} = \frac{2\sqrt{2}}{3}\delta^{3/2} + O(\delta^{5/2}), \tag{S37}$$

where $\rho_{1,\perp}, \rho_{2,\perp}, \rho_{3,\perp}, \rho_{4,\perp}$ correspond to the four eigenstates characterized by $m = -\frac{3}{2}, -\frac{1}{2}, \frac{1}{2}, \frac{3}{2}$.

At the point $\vec{a} + \vec{d}_\parallel = (\cos\theta + \delta\sin\theta, \sin\theta - \delta\cos\theta)$, the Hamiltonian takes the form
$$H_\parallel^{3/2}(\theta) = (\cos\theta + \delta\sin\theta)s_x^{3/2} - is_y^{3/2} + (\sin\theta - \delta\cos\theta)\sin\theta\, s_z^{3/2}. \tag{S38}$$

The eigenvalues for $H_\parallel^{3/2}(\theta)$ are $E_\parallel = \frac{2}{3}m\delta$ where $m = -\frac{3}{2}, -\frac{1}{2}, \frac{1}{2}, \frac{3}{2}$. And the phase rigidities for $H_\parallel^{3/2}(\theta)$ are

$$\rho_{1,\parallel} = \rho_{4,\parallel} = \delta^3 + O(\delta^5), \tag{S39}$$
$$\rho_{2,\parallel} = \rho_{3,\parallel} = \frac{\delta^3}{3} + O(\delta^5). \tag{S40}$$

In summary, we have $\rho_{i,\perp} \propto |\delta|^{3/2}$ and $\rho_{i,\parallel} \propto |\delta|^3$, which cover the $\delta < 0$ scenarios.

### IX. Nth order EPs with asymptotic form $E \propto \Delta^{1/N}$ near an EP

First we rewrite the real asymmetric hopping Hamiltonian $H$ in Eq. (1),

$$H = \begin{pmatrix} 0 & t_1 & 0 & 0 & 0 \\ t_2 & 0 & t_1 & 0 & 0 \\ 0 & t_2 & 0 & \ddots & 0 \\ 0 & 0 & \ddots & \ddots & t_1 \\ 0 & 0 & 0 & t_2 & 0 \end{pmatrix}, \quad (S41)$$

where $t_1 = t - 1$ and $t_2 = t + 1$. The eigenvalues are $E = 2\sqrt{t^2 - 1}\cos\left(\frac{m\pi}{N+1}\right), m = 1, 2, \cdots, N$. Near the EPs at $t = \pm 1$, the eigenvalues are proportional to $\Delta^{1/2}$. In the discussion part in the main text, we mentioned that the conventional order-N EP associated with $\Delta^{1/N}$ type eigenvalues also exist in the context of asymmetric hopping model. Suppose we take $t = 1$, we have

$$H_{EP} = \begin{pmatrix} 0 & 0 & 0 & 0 & 0 \\ 2 & 0 & 0 & 0 & 0 \\ 0 & 2 & 0 & \ddots & 0 \\ 0 & 0 & \ddots & \ddots & 0 \\ 0 & 0 & 0 & 2 & 0 \end{pmatrix}, \quad (S42)$$

which obviously corresponding to an order-N EP. Introducing $\Delta$ with $|\Delta| = 0^+$ into the upper right corner of $H_{EP}$, we have

$$H_1 = \begin{pmatrix} 0 & 0 & 0 & 0 & \Delta \\ 2 & 0 & 0 & 0 & 0 \\ 0 & 2 & 0 & \ddots & 0 \\ 0 & 0 & \ddots & \ddots & 0 \\ 0 & 0 & 0 & 2 & 0 \end{pmatrix}. \quad (S43)$$

The eigenvalues of $H_1$ take the form $E_i = a_i \Delta^{1/N}, i = 1, \cdots, N$ when $|\Delta| = 0^+$.

Following similar procedures as in Section I, we can easily derive the left eigenvector and right eigenvector for any eigenstate $E_j = a_j \Delta^{1/N}$ to be

$$|\psi_j^R(\Delta)\rangle = \left(A_{N-1}\Delta^{\frac{N-1}{N}}, A_{N-2}\Delta^{\frac{N-2}{N}}, \cdots, A_1\Delta^{\frac{1}{N}}, 1\right)^T,$$

$$\langle\psi_j^L(\Delta)| = \left(1, B_1\Delta^{\frac{1}{N}}, \cdots, B_{N-2}\Delta^{\frac{N-2}{N}}, B_{N-1}\Delta^{\frac{N-1}{N}}\right),$$

where $A_i$ and $B_i$ are constants. The phase rigidity for the $E_j$ state is then

$$\rho_j \propto \Delta^{\frac{N-1}{N}}. \quad (S44)$$

That is, the exponent of phase rigidity is $\frac{N-1}{N}$, which is different from the $\sqrt{\Delta}$ type EPNs, which are associated with exponent of $\frac{N-1}{2}$.

We note that all exponents of $\frac{N-1}{n}$ with $1 \leq n \leq N$ can be found by introducing perturbation to $H_{EP}$. Adding a $\Delta$ to an arbitrary $(i, i+n-1)$-th entry of $H_{EP}$ where $n = 2, 3, \cdots, N$, the eigenvalues near the EPN would take the form

$$E_j = a_j \Delta^{\frac{1}{n}}, \quad (S45)$$

with possibly some eigenvalues remaining zero. When $n = 2$, the $\sqrt{\Delta}$ type order-N EP is recovered. And when $n = N$, the aforementioned $\sqrt[N]{\Delta}$ type order-N EP is then recovered. The phase rigidity for the $E_j$ state is then $\rho_j \propto \Delta^{\frac{N-1}{n}}$, which can be easily derived similarly.

To be specific, we take $N = 4$ as an example. The Hamiltonian $H_{EP}$ is

$$H_{EP}^{N=4} = \begin{pmatrix} 0 & 0 & 0 & 0 \\ 2 & 0 & 0 & 0 \\ 0 & 2 & 0 & 0 \\ 0 & 0 & 2 & 0 \end{pmatrix}. \tag{S46}$$

The results for the different perturbations are summarized as follows:

(1) Introducing a perturbation so that $(H_{EP}^{N=4})_{i,i} = \Delta$ where $i = 1,2,\cdots,4$, one eigenvalue turns $E = \Delta$ and other eigenvalues remain zero. The phase rigidity for the $E = \Delta$ eigenstate is $\rho \sim \Delta^3$, which is consistent with the $\rho \sim |\epsilon|^{N-1}$ rule in the main text.

(2) Introducing a perturbation so that $(H_{EP}^{N=4})_{1,2} = \Delta$ or/and $(H_{EP}^{N=4})_{2,3} = \Delta$ or/and $(H_{EP}^{N=4})_{3,4} = \Delta$, the eigenvalues take the form $E \sim \Delta^{1/2}$ except that some eigenvalues may remain zero. The phase rigidities for the eigenstates are $\rho \sim \Delta^{3/2}$.

(3) Introducing a perturbation so that $(H_{EP}^{N=4})_{1,3} = \Delta$ or/and $(H_{EP}^{N=4})_{2,4} = \Delta$, the eigenvalues take the form $E \sim \Delta^{1/3}$ except that some eigenvalues may remain zero. The phase rigidities for the eigenstates are $\rho \sim \Delta$.

(4) Introducing a perturbation so that $(H_{EP}^{N=4})_{1,4} = \Delta$, the eigenvalues take the form $E \sim \Delta^{1/4}$. The phase rigidities for the eigenstates are $\rho \sim \Delta^{3/4}$.